\documentclass[prl,twocolumn]{revtex4-1}
\usepackage{multirow}
\usepackage{graphicx}
\usepackage{amsmath}
\usepackage{enumerate}
\usepackage{epstopdf}
\usepackage{times}
\usepackage{color}
\usepackage{bm}
\usepackage{comment}

\DeclareMathAlphabet      {\mathbfit}{OML}{cmm}{b}{it}

\begin{document}


\title{Fluctuation-Driven Selection at Criticality in a Frustrated Magnetic System: \\
the Case of Multiple-${\bm k}$ Partial Order on the Pyrochlore Lattice}


\author{Behnam Javanparast$^1$, Zhihao Hao$^1$, Matthew Enjalran$^2$, Michel J. P. Gingras$^{1,3,4}$}
\address{\footnotesize{$^1$Department of Physics and Astronomy, University of Waterloo, Waterloo, Ontario, N2L 3G1, Canada\\
$^2$Physics Department, Southern Connecticut State University, 501 Crescent Street, New Haven, CT 06515-1355, USA\\
$^3$Canadian Institute for Advanced Research, 180 Dundas St. W., Toronto, Ontario, M5G 1Z8, Canada\\
$^4$Perimeter Institute for Theoretical Physics, 31 Caroline St. N., Waterloo, ON, N2L 2Y5, Canada}}

\date{\rm\today}

\begin{abstract}

We study the problem of partially ordered phases with periodically arranged disordered (paramagnetic) sites on the pyrochlore lattice, a network of 
corner-sharing tetrahedra. The periodicity of these phases is characterized by one or more wave vectors 
${\bm k} = \{\frac{1}{2}\frac{1}{2}\frac{1}{2}\}$. Starting from a general microscopic Hamiltonian including anisotropic nearest-neighbor 
exchange, long-range dipolar interactions and second- and third-nearest neighbor exchange, we identify using standard mean-field theory (s-MFT)
an extended range of interaction parameters that support partially ordered phases.
  We demonstrate that thermal fluctuations ignored in s-MFT
are responsible for the selection of one particular partially ordered phase, e.g. the ``4-${\bm k}$'' phase over the ``1-${\bm k}$'' phase. 
We suggest that the transition into the 4-${\bm k}$
phase is continuous with its critical properties controlled by the cubic fixed point of a Ginzburg-Landau theory with
a 4-component vector order-parameter. 
By combining an extension of the Thouless-Anderson-Palmer method originally used to study fluctuations in spin glasses
with  parallel-tempering Monte-Carlo simulations, we establish the phase diagram for different types of  partially ordered phases. 
Our results elucidate  the long-standing puzzle concerning the origin of the  4-${\bm k}$ 
partially ordered phase observed in the Gd$_2$Ti$_2$O$_7$ dipolar pyrochlore antiferromagnet
below its paramagnetic phase transition temperature.

\end{abstract}


\maketitle


 \begin{figure}
\begin{center}
\includegraphics[width=9.5 cm]{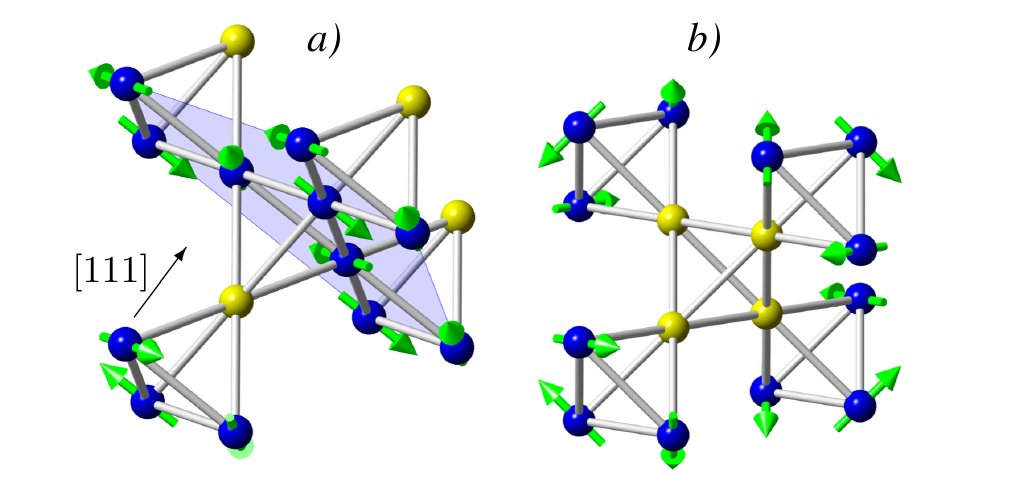}
\caption{(Color online) 
 The yellow (arrowless) sites are disordered (paramagnetic).
{\it a)} a 1-${\bm k}$ state with ${\bm k}$ along $[111]$.
The spins on the (blue)  ordered sites form a $120^\circ$ pattern on a triangle that does not share a corner
with a disordered site. These ordered sites form
a kagome plane (blue-shaded plane) perpendicular to the $[111]$ direction.
 {\it b)} 4-${\bm k}$ state arising from the superposition of four 1-$\bm k$ states.
\label{fig:1k4k}}
\end{center}
\end{figure}

Highly frustrated magnetism is one of the paradigms of modern condensed matter physics
 \cite{Frustrated_Magnetism}.
In frustrated magnets, the combination of lattice geometry and competing interactions often leads to 
degenerate classical states. The degeneracies are generally accidental as they are not
protected by the symmetries of the spin Hamiltonian.
Yet, the degenerate states may be related by transformations that form an {\it emergent} symmetry group. 
Near a continuous phase transition, these approximate symmetries provide ``organizing principles''
in determining the critical properties by distinguishing relevant perturbations from irrelevant ones.
In the most interesting case, the leading degeneracy-lifting perturbations, 
which may  be relevant or irrelevant in the renormalization group sense, 
are thermal or quantum fluctuations -- a phenomenon called 
order-by-disorder (ObD)  \cite{Villain_ObD,Henley_ObD,Yildirim_ObD, shender_ObD}.
The competition among diverse degeneracy-lifting
effects can result in a modulated long-range ordered state
at nonzero wave vector ${\bm k}$, which may  or may not be commensurate with the lattice \cite{Jensen,Rossat,Incomm_DyY,Reimers,Chatto,Incomm_1,Incomm_2}.
In some cases, a number of superposed symmetry-related ${\bm k}$ modes within the first Brillouin zone
 form a so-called {\it multiple-${\bm k}$ order} \cite{Jensen,Rossat,Mul-k_Nd,Mul-k_pyro}.
 A particular interesting form of such modulated magnetism is
a partially ordered state (POS) with periodically arranged ``paramagnetic'' 
sites \cite{POS_CsCoBr3,POS_CuFeO2,POS_UNi4B,POS_Sr2YRuO6}.
These fluctuating magnetic moments decimate a fraction of the energy-costly frustrated bonds while retaining an extensive entropy, 
hence lowering the free energy.

In this Letter, we study the convergence of  the aforementioned phenomena
(emergent symmetry, multiple-${\bm k}$ POS and fluctuation-induced degeneracy-lifting)
in an extensively studied class of frustrated magnetic materials, 
the insulating $R_2M_2$O$_7$ pyrochlore oxides \cite{Gardner_RMP}.
 In these, the $R^{3+}$ magnetic rare-earth ions (e.g. Gd, Tb, Er, Yb)
occupy the vertices of a network of corner-sharing tetrahedra -- the ``pyrochlore'' lattice (see Fig. 1).
$M^{4+}$ ($M$=Ti, Sn, Zr, Ge) is non-magnetic.
The competition between four types of nearest-neighbor anisotropic interactions
and the nature of the single-ion magnetic anisotropy are 
largely responsible for the wealth of phenomena displayed by the $R_2M_2$O$_7$ materials \cite{Gardner_RMP}.
In this paper, we focus on a general description for the perplexing yet rich physics of multi-${\bm k}$ 
partial magnetic ordering in pyrochlore oxides and not on any material-specific issues.  
  

%

A multi-${\bm k}$ POS (see Fig. 1b) is believed to exist
in  Gd$_2$Ti$_2$O$_7$ \cite{stew}  for temperature 0.7 K $ \lesssim T \lesssim T_c\sim 1$ K in which the
POS is a superposition of spin density waves with wave vectors ${\bm k}=\{ \frac{1}{2} \frac{1}{2} \frac{1}{2}\}$
\cite{stew,Raju_Gd2Ti2O7,Champion_Gd2Ti2O7,Matt1,cepas1,cepas2,Wills_Gd2Sn2O7}.
While this compound has been the subject of a number of investigations 
\cite{Raju_Gd2Ti2O7,Champion_Gd2Ti2O7,Ramirez_Gd2Ti2O7,Bonville_Gd2Ti2O7,
stew,Petrenko_Gd2Ti2O7,Sosin_magnetocaloric,Sosin_adiab,Yaouanc_Gd2Ti2O7_DOS,
Reotier_GTOcrit,Dunsiger_Gd2Ti2O7,Liao_Gd2Ti2O7,Petrenko_100},
the mechanism responsible for the selection of 4-${\bm k}$ order has not been identified.
Further, after fifteen years of  research on the Tb$_2$Ti$_2$O$_7$ spin liquid candidate \cite{Gardner_Tb2Ti2O7},
evidence has recently begun accumulating that short-range magnetic correlations develop below $T \sim 0.4$ K 
in the form of broad elastic neutron intensities at  ${\bm k}=\{ \frac{1}{2} \frac{1}{2} \frac{1}{2} \}$
\cite{Fennell_Tb2Ti2O7,Petit_Tb2Ti2O7,Fritsch_Tb2Ti2O7,Taniguchi_Tb2Ti2O7,Guitteny_Tb2Ti2O7}.
No theory has explained the origin of these correlations.
As Tb$_2$Ti$_2$O$_7$ and Gd$_2$Ti$_2$O$_7$ constitute two out of the six magnetic $R_2$Ti$_2$O$_7$ pyrochlore
compounds that exist ($R$=Gd, Tb, Dy, Ho, Er and Yb),  it may be that
${\bm k}=(\frac{1}{2} \frac{1}{2} \frac{1}{2})$ order is not 
unusual among the plethora of $R_2M_2$O$_7$  materials,
or even  $AR_2B_4$ rare-earth spinels 
($A$=Cd, Mg; $B$=S, Se and $R$ is a rare-earth) \cite{Lau_spinels,Lago_spinel,Wong}.

In our work, we identify an extensive range of exchange parameters 
able to support  ${\bm k}=(\frac{1}{2}\frac{1}{2}\frac{1}{2})$ partial order through a standard mean-field theory (s-MFT) study.  
The s-MFT free energy displays,
up to quartic order in the order parameters, an {\it  emergent} $O(4)$ symmetry.
The transition into the POS is thus described by the Ginzburg-Landau free-energy, $\cal{F}$, of an $n$-component vector model ($n=4$) \cite{cond,Pelissetto_review}.  
The most relevant perturbation is a ``cubic anisotropy'' which  breaks the $O(4)$ symmetry.
We thus identify the physical origin of cubic anisotropy as  thermal fluctuations
{\it  beyond} s-MFT. 
The $O(4)$ symmetry-breaking and 
selection of either 1-${\bm k}$ or 4-${\bm k}$ order in this model is an example of thermal ObD. 
We stress, however,  that our problem conceptually departs significantly from the more common cases of ObD where 
``small'' thermal or quantum fluctuations are typically considered. 
This is
valid for temperatures much lower 
than the critical temperature where the harmonic approximation is often justified. 
For example, in Gd$_2$Ti$_2$O$_7$, where another transition occurs at $T \sim 0.7$ K $< T_c$ 
\cite{Ramirez_Gd2Ti2O7,Yaouanc_Gd2Ti2O7_DOS,Reotier_GTOcrit,Liao_Gd2Ti2O7,Petrenko_100}, 
the low temperature fluctuations have no bearing on the state-selection at $T_c$.



Given the above considerations, we develop an extension of the Thouless-Anderson-Palmer (TAP) 
method, E-TAP, to obtain the phase diagram of 1-${\bm k}$ and 4-${\bm k}$ orders.
We use Monte Carlo simulations to confirm the E-TAP predictions that
1-${\bm k}$ and 4-${\bm k}$ POSs are selected in  different portions of the phase diagram.
 We suggest that the phase transition at $T_c \sim 1$ K in Gd$_2$Ti$_2$O$_7$ belongs to the 
above  $n=4$ cubic universality class. From a broader methodological perspective,  
our E-TAP method for spin models with anisotropic interactions could be applied to 
other problems in frustrated magnetism where the question of state selection at $T_c$ is of interest.

{\it Model} -- We consider the general Hamiltonian, 
${\cal{H}}={\cal{H}}_{0}+{\cal{H}}_{\mathrm{dip}}+ {\cal{H}}_2+ {\cal{H}}_3$, 
for classical spins, ${\bm S}_i$, on the pyrochlore lattice:
\begin{subequations}
\label{eq:hami}
\begin{eqnarray}
&&{\cal{H}}_0\equiv\sum_{\langle i,j\rangle}[J\,\bm{S}_i\cdot\bm{S}_j+J_{\mathrm{DM}}\,\hat{d}_{ij}\cdot(\bm{S}_i\times\bm{S}_j)
+
\nonumber\\
&&J_{\mathrm{Ising}}\,(\bm{S}_i\cdot \hat{z}_i)(\bm{S}_j\cdot \hat{z}_j)+J_{\mathrm{pd}} \,S_{i}^{\alpha}\Lambda^{(\alpha\beta)}_{ij}S^{\beta}_{j}].\\
&&{\cal{H}}_{\mathrm{dip}}=J_{\mathrm{dip}}\sum_{i>j}S_{i}^{\alpha}    \Lambda^{(\alpha\beta)}_{ij} S_{j}^{\beta} . \\
&&{\cal{H}}_2\equiv\sum_{\langle\langle ij\rangle\rangle}J_2\,\bm{S}_i\cdot \bm{S}_j , \;\;
{\cal{H}}_3\equiv\sum_{\langle\langle\langle ij\rangle\rangle\rangle}J_3\,\bm{S}_i\cdot\bm{S}_j.
\end{eqnarray}
\end{subequations}
Here,  ${\cal{H}}_{0}$ includes all possible symmetry-allowed nearest-neighbor (n.n.) bilinear interactions: isotropic ($J$), 
Dzyaloshinskii-Moriya (DM) ($J_{\mathrm{DM}}$), Ising  ($J_{\mathrm{Ising}}$) and pseudo-dipolar 
($J_{\mathrm{pd}}$) \cite{Thompson_Yb2Ti2O7,McClarty_Er2Ti2O7}. 
Of exchange origin, all these couplings can be positive or negative.
 Unit vectors $\hat{d}_{ij}$ are chosen such that positive and negative $J_{\mathrm{DM}}$ 
correspond to direct and indirect DM interactions, respectively \cite{Elhajal2005}. $\hat{z}_i$ is the local cubic $[111]$ direction 
at site $i$ and 
$\Lambda_{ij}^{(\alpha\beta)}\equiv (\delta_{\alpha\beta}/r_{ij}^3-3r_{ij;\alpha}r_{ij;\beta}/r_{ij}^5)$.
${\cal{H}}_{\mathrm{dip}}$ is the long-range magnetostatic dipole-dipole interaction with strength 
$J_{\mathrm{dip}}=\frac{\mu_0\mu^2}{4\pi r_{\mathrm{nn}}^3}$, where $r_{\mathrm{nn}}$ 
is the nearest-neighbor distance.
$r_{ij}$ is measured in units of $r_{\mathrm{nn}}$  and $\mu$ is the magnetic moment.
Our convention for ${\cal{H}}_0$ \cite{Thompson_Yb2Ti2O7,McClarty_Er2Ti2O7} 
differs from that used by other groups \cite{Curnoe2008,balents}
but the two are related by a linear-transformation.
${\cal{H}}_2$ and ${\cal{H}}_3$ are second and third n.n. exchange interactions, respectively.

We begin by studying  ${\cal{H}}_0$ using s-MFT \cite{Reimers,suppl,Enjalran_MFT,Matt1}. 
For a large region of parameter space,  
a {\it degenerate} line of modes with momenta $\{hhh\}$ first becomes critical at $T_c$ \cite{Raju_Gd2Ti2O7,Matt1,cepas1,cepas2}.
s-MFT calculations show that ${\cal{H}}_{\mathrm{dip}}$ lifts the degeneracy by weakly selecting soft modes at the four 
 $\bm{L} \equiv  \{\frac{1}{2}\frac{1}{2}\frac{1}{2} \}$ 
points within the first Brillouin zone, which we label $\bm{k}_{a}$ ($a=0,1,2,3$) \cite{cepas1,Matt1}. 
Other perturbations to ${\cal{H}}_0$, such as ${\cal{H}}_2$ and ${\cal{H}}_3$, 
can have similar effects \cite{suppl,Matt1,cepas1,cepas2,Wills_Gd2Sn2O7}.
Here, we focus on ${\cal{H}}_{\mathrm{dip}}$ since it can be of prominence in  rare-earth pyrochlore oxides \cite{Gardner_RMP}. 
In particular, we choose the Gd$_2$Ti$_2$O$_7$  value  $J_{\mathrm{dip}}/J \sim 0.2$ \cite{Raju_Gd2Ti2O7}
as an examplar.
For completeness, and of possible relevance to magnetic pyrochlores with ions having a small magnetic moment, 
we present some results in the Supplemental Material (Sup. Mat.) \cite{suppl} 
that use the notation of Ref.~\cite{balents} for ${\cal H}_0$ and 
where the $\{hhh\}$ degeneracy-lifting originates from ${\cal H}_2$ as opposed to ${\cal H}_{\rm dip}$.

The direct space spin configurations corresponding to a 1-$\bm k$ state, with
$\bm{k}= \bm{k}_0 \equiv (\frac{1}{2}\frac{1}{2}\frac{1}{2})$, is illustrated in Fig. \ref{fig:1k4k}a.
Denoting the corresponding spin direction at site $\bm{r}_i$ as $\hat{e}_{\bm{k}_0}(\bm{r}_i)$, 
with $\hat{e}_{\bm{k}_a}(\bm{r}_i)$ ($a=1,2,3$) defined similarly \cite{suppl}, we 
introduce the order parameter $\psi_a$ for a particular ordering wave vector $\bm{k}_a$ as
$\psi_{a}=  {\frac{1}{N}}\sum_{i}\bm{S}_i\cdot\hat{e}_{\bm{k}_{a}}(\bm{r}_i)$, where 
the summation goes over all $N$ sites 
of the pyrochlore lattice and $\bm{\psi}\equiv(\psi_0, \cdots, \psi_3)$
defines a 4-dimensional vector.


For a wide range of $(J,J_{\rm DM},J_{\rm Ising},J_{\rm pd},J_{\rm dip},J_2,J_3)$ couplings,
s-MFT predicts a second order transition with ordering at momenta $\bm{k}_a$.
Interestingly, we find that up to quartic order in the $\{ \psi_a \}$ order parameters,
the s-MFT free-energy displays an {\it emergent} $O(4)$ symmetry \cite{suppl}.
While the $O(4)$ symmetry is spoiled at higher order in $\{ \psi_a \}$, 
{\it no}  selection of 1-${\bm k}$ (which has only one $\psi_a \ne 0$)
vs 4-${\bm k}$ (which has four $\psi_a$ with the same nonzero amplitude)
occurs {\it at any} order \cite{suppl}.

The observation of an emergent $O(4)$ symmetry at quartic order in s-MFT helps us to recognize that 
a ``cubic anisotropy'' in the $\{ \psi_a \}$ is the most relevant symmetry-allowed perturbation.
We then write the  Ginzburg-Landau free energy ${\cal F}(\{ \psi_a \})$  \cite{suppl} of the system,
up to quartic order in the $\{\psi_a\}$, that includes a cubic anisotropy term ($v\ne 0$):
\begin{equation}\label{eq:freeenergy}
{\cal{F}}=\frac{1}{2}r\sum_{a=0}^{3}\psi_{a}^2+u\left(\sum_{a=0}^{3}\psi_{a}^2\right)^2+v\sum_{a=0}^{3}\psi_{a}^4  .
\end{equation}
Eq. \eqref{eq:freeenergy} is the celebrated $n$-vector model ($n=4$) with cubic anisotropy \cite{cond} with $r=a_0(T-T_c)$ and $a_0>0$.
This model, a cornerstone of the theory of critical phenomena \cite{cond,Pelissetto_review},
has been extensively investigated with numerous methods  \cite{Ferer1981,Newman1982, Caselle1998, carmona2000}.
In $3D$, for $n>n_c\approx 2.89$ and $u>0$ \cite{carmona2000}, the model undergoes a second-order  transition 
into a phase where all $\psi_a$ have the same amplitude if $v>0$. 
The universality class is  controlled by the cubic
fixed point \cite{cond} with distinct critical exponents from those of the isotropic $O(n)$ fixed point \cite{Pelissetto_review,carmona2000}. 
For $v<0$, the phase transition is a fluctuation-induced first-order transition to a state with only one nonzero $\psi_a $.
Therefore, we conclude that for $v>0$ the magnetic order is defined by a superposition of four ${\bm k}_a$ spin states
(4-${\bm k}$ state) while for $v<0$ the order is defined by a single ${\bm k}_a$ spin structure (1-${\bm k}$ state). 
 Both  states are POSs since the spins on $1/4$ of the sites remain disordered 
(see Fig. \ref{fig:1k4k}).  
As stated above, s-MFT predicts that 1-${\bm k}$ and 4-${\bm k}$ have the same $\cal{F}$ (i.e. $v=0$). 
We thus identify thermal {\it fluctuations} as the 
mechanism for generating $v\ne 0$ leading to a selection of 1-$\bm{k}$ vs 4-$\bm{k}$.
To expose how fluctuations may lead 
{\it at the microscopic level} to a selection {\it directly at} $T_c$,
we devise and then use an E-TAP method \cite{TAPmain,Yadidah,Plefka}.

{\it Extended TAP Method} -- 
A magnetic moment at a particular lattice site experiences a local field due to its neighbors. 
At the s-MFT level, the presence of the spin at the site of interest affects its local field indirectly. 
This is an artifact of s-MFT. The Onsager reaction field (ORF) introduces a term that cancels this unphysical effect.
The TAP approach provides a systematic way to implement the ORF correction \cite{naiveTAP}. 
In this work, we devise an extension of  the TAP method, E-TAP,  
where the ORF is the first term of a series originating from nonzero on-site fluctuations,  e.g. 
$ \langle S^{\alpha}_{i} S^{\beta}_{i}\rangle \neq \langle S^{\alpha}_{i} \rangle \langle S^{\beta}_{i}\rangle$ \cite{suppl, Plefka, Yadidah}.



To compute the E-TAP corrections, we consider a perturbative expansion of the Gibbs free-energy, $G$,
in inverse temperature $\beta$ \cite{suppl,Yadidah}. 
 \begin{equation}
 \label{gibbs}
\beta G=-\ln \Big( \text{Tr}[\exp\big(-\beta {\cal{H}}+ \sum_{i} \bm \lambda_{i}\cdot (\bm S_{i}-\bm m_{i})\big)] \Big) .
 \end{equation}
Here $\bm m_{i}$ is the local magnetization and the $\bm \lambda_{i}$ vector
is a Lagrange multiplier; different from the local 
mean field at site $i$ by a factor of $\beta$ \cite{suppl,Yadidah}.  
Defining $ \tilde{G}(\beta)\equiv \beta G$, the first and second terms in the expansion,
 $\tilde{G}(0)/\beta$ and $\tilde{G}^{\prime}(0)$, 
where the prime represents differentiation with respect to $\beta$, 
are the s-MFT entropy and energy, respectively. 
The third term, $\Omega \equiv \tilde{G}^{\prime\prime}(0)\beta/2$, 
is the first correction beyond s-MFT, arising from fluctuations \cite{suppl,Yadidah}:
\begin{equation}
\Omega = -\frac{\beta}{4}\sum_{{i},{j}} 
\sum_{\alpha\beta\gamma\delta} J^{\alpha\gamma}_{{i}{j}} J^{\beta\delta}_{{i}{j}} \chi^{\alpha\beta}_{i}  \chi^{\gamma\delta}_{j} .
\label{TAP2}
\end{equation}
Here,  $ \chi_{i}^{\alpha\beta}=\langle S^{\alpha}_{i}S^{\beta}_{i}\rangle-\langle S^{\alpha}_{i} \rangle \langle S^{\beta}_{i}\rangle$ is the on-site susceptibility. 
Since $\chi_i^{\alpha\beta}$ is quadratic in  $\{ \psi_a \}$ \cite{suppl},
$\Omega$ in Eq. (\ref{TAP2}) is therefore of {\it quartic order} in $\{\psi_a\}$.
The first order E-TAP correction, $\Omega$, can thus, in principle, generate
a finite cubic anisotropy term in $\cal{F}$ and select 1-${\bm k}$  or 
4-${\bm k}$ depending
on the bilinear spin-spin interaction matrix $J_{ij}^{\alpha\beta}$
 defined through ${\cal{H}} \equiv \sum_{(i>j);\alpha,\beta}  J_{ij}^{\alpha\beta} S_i^\alpha S_j^\beta$ in  Eq. (\ref{eq:hami}).
 
\begin{figure}
 \begin{center}
\includegraphics[width=8.5cm]{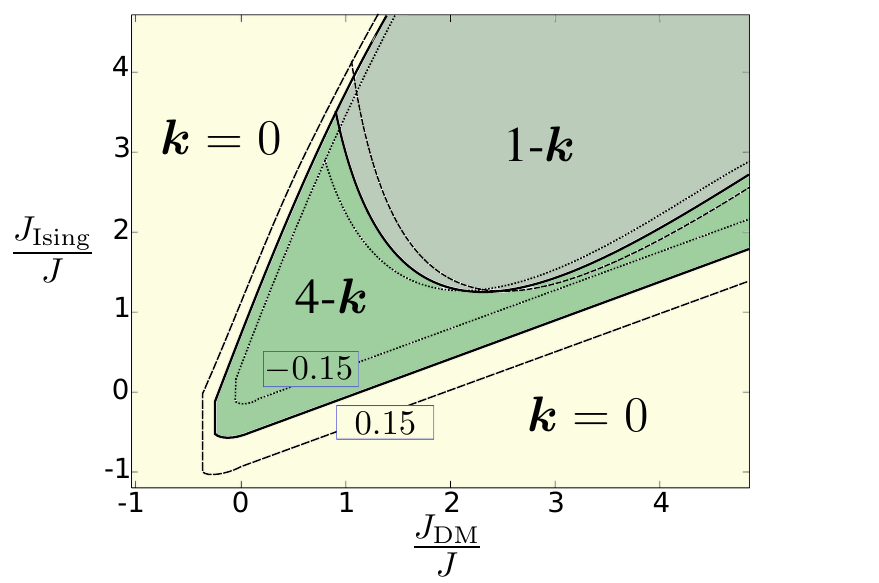}
\caption{ (Color online) Ordering wave vectors at $T=T_c$ obtained 
from s-MFT. The combined area denoted 1-${\bm k}$ and 4-${\bm k}$  displays
$(\frac{1}{2}\frac{1}{2}\frac{1}{2})$ order but 
1-${\bm k}$ and 4-${\bm k}$  are degenerate at the s-MFT level. The first order E-TAP correction 
Eq. (\ref{TAP2}) applied in this regime selects either 1-${\bm k}$ or 4-${\bm k}$.
The ${\bm k}=0$ region encompass all states for which all sites are fully ordered and each primitive 4-site tetrahedron basis
has the same local spin order.
The dashed (0.15) and dotted (-0.15) contours mark the boundaries for the corresponding
$J_{\rm pd}$ values.
}
\label{fig:Wedge}
\end{center}
\end{figure}

We calculate  $\Omega$ \cite{suppl} for the region in parameter space with 
$(\frac{1}{2} \frac{1}{2} \frac{1}{2})$ ordering
(the dark-shaded 1-${\bm k}$/4-${\bm k}$ wedge in Fig. \ref{fig:Wedge}).
Specifically, we compute  
 $\delta\Omega \equiv \Omega_{1{\bm k}} - \Omega_{4{\bm k}}$.
$\delta\Omega < 0$ indicates a 1-${\bm k}$ selection and conversely for $\delta\Omega > 0$, 
with the $\Omega$ contribution splitting the wedge into 1-${\bm k}$ and 4-${\bm k}$ sectors.
In particular, for a dipolar Heisenberg model with
$J_{\rm Ising}=J_{\rm DM}=J_{\rm pd}=0$, relevant to Gd$_2$Ti$_2$O$_7$ \cite{Raju_Gd2Ti2O7}, E-TAP calculations
predict a 4-${\bm k}$ state selection at $T_c$, as observed in this compound \cite{stew}.


{\it Monte Carlo Simulations} -- 
We performed parallel tempering classical Monte Carlo simulations to check the E-TAP predictions.
We pick, somewhat arbitrarily, two sets of interaction parameters 
corresponding to the 1-${\bm k}$ and 4-${\bm k}$ regions in Fig. \ref{fig:Wedge}.
One set  is the simplest model for a spin-only 
$^8S_{7/2}$ ($L=0$, $S=7/2$, $J=L+S$) state for Gd$^{3+}$ in Gd$_2$Ti$_2$O$_7$, with $J_{\rm dip}/J=0.18$ \cite{Raju_Gd2Ti2O7},
$J_{\rm Ising}=J_{\rm DM}=J_{\rm pd}=0$. 
We  include, as in a previous Monte Carlo work \cite{cepas2}, weak 
ferromagnetic second n.n. interaction (e.g.  $-0.02J\lesssim J_2 < 0$)  to stabilize a temperature range
wide enough to numerically resolve $(\frac{1}{2}\frac{1}{2}\frac{1}{2})$ order \cite{suppl,cepas2}.
We measure the magnitude of $\bm{\psi}$:
\begin{equation}
|\bm{\psi}|^2=\left\langle\sum_{a=0}^3\psi_a^2\right\rangle	,
\end{equation}
where $\langle \ldots \rangle $ denotes a thermal average.

In addition to its magnitude, we identify the orientation of $\bm{\psi}$ at $T < T_c$ 
through two additional order parameters, $d_{i\bm k}$ ($i=1,4$) defined below. 
On the surface of a four-dimensional unit hyper-sphere, 
there are eight ``1-${\bm k}$ points'' corresponding to 1-${\bm k}$ states including $(\pm 1,0,0,0)$,
$(0,\pm 1,0,0)$, etc.
Similarly, there are sixteen `` 4-${\bm k}$ points'' on the hyper-sphere \cite{suppl}. 
Given a certain spin configuration, we calculate $\hat{\psi}\equiv \bm{\psi}/|\bm{\psi}|$. 
The $d_{i\bm k}$ ($i=1,4$) are then defined as the {\it minimum} Euclidean distance 
between point $\hat{\psi}$ and all of the $i$-${\bm k}$ points.
 The thermal average of $d_{1\bm k}$ ($d_{4\bm k}$) is expected to decrease if the system enters a
1-${\bm k}$ (4-${\bm k}$) state for $T<T_c$. 

\begin{figure}
 \begin{center}
\includegraphics[width=9.5cm]{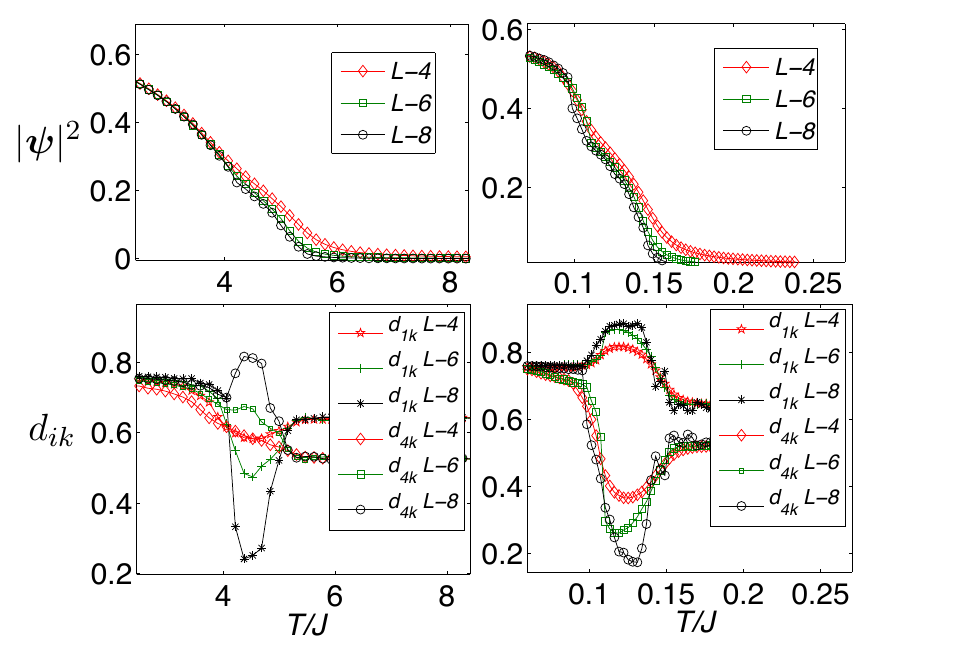}
\caption{(Color online) Monte Carlo simulations results. 
{\it Top row:}  growth of $\vert {\bm{\psi}} \vert ^2$. 
{\it Bottom row:} 
value of $d_{i{\bm k}}$ as a function of $T/J$. {\bf Left column:} 
$J_{\mathrm{pd}}=0$, ${J_{\mathrm{dip}}}/{J}=0.18$, ${J_{2}}/{J}=-1$, ${J_{\mathrm{Ising}}}/{J}=10$, 
${J_{\mathrm{DM}}}/{J}=10$ (direct DM) and ${J_3}/{J}=1$ (a point in 1-$\bm k$ region).  {\bf Right column:} 
$J_{\mathrm{pd}}=J_{\mathrm{DM}}=J_{\mathrm{Ising}}=0$, 
${J_{\mathrm{dip}}}/{J}=0.18$ and ${J_2}/{J}=-0.02$ (a point in 4-$\bm k$ region). 
Note: The $d_{i{\bm k}}$ are nonzero in the paramagnetic phase because
their value is equal to the average distance of a random point on a 4-dimensional 
hypersphere from the 1-$\bm k$ and 4-$\bm k$ points.
} 
\label{fig:MC}
\end{center}
\end{figure}

The results for two sets of interaction parameters in the 1-$\bm{k}$ and 4-$\bm{k}$
 regions of Fig. \ref{fig:Wedge} are shown in the left  and right columns of Fig. \ref{fig:MC}, respectively.
The growth of $|\bm{\psi}|^2$ at $T_c/J\sim 5.5$ and $T_c/J \sim 0.15$
shows that the system orders with ${\bm k}=(\frac{1}{2}\frac{1}{2}\frac{1}{2})$. 
In the left column,  the 1-${\bm k}$ state is selected at $T_c$, as
 indicated by a minimum for $d_{1\bm{k}}$ and a maximum for $d_{4\bm{k}}$.
The system orders in a 4-${\bm k}$ state in the right column.
The separation of $d_{1\bm k}$ and $d_{4\bm k}$ for both cases in Fig. \ref{fig:MC} 
accentuates as the linear dimension $L$ of the system increases,
indicating that the selection of either 1-${\bm k}$ or 4-${\bm k}$ survives in the thermodynamical limit.  
These results are consistent with the predictions from the E-TAP calculations at $T_c$.
Unfortunately, the large computational resources required for simulations 
with long-range dipolar interactions prevent us from investigating the order of the phase transitions in Fig. \ref{fig:MC}.

The kinks in $|\bm{\psi}|^2$ and merging of $d_{1{\bm k}}$ and $d_{4{\bm k}}$
 indicate  the system enters into a distinct phase at $T/J \lesssim 4$ and $T/J 
\lesssim 0.1$ in the left and right columns of Fig. \ref{fig:MC}. 
Since in a 2-${\bm k}$ state $d_{1k}\equiv d_{4k}=\sqrt{2-\sqrt{2}}\simeq 0.765$ \cite{suppl}, 
the results for $d_{i{\bm k}}$ suggest that the low-temperature region may be a 2-$\bm k$ state.
We note that the latest single-crystal neutron diffraction results \cite{Stewart_comm}
indicate that the low-temperature state ($T<0.7$ K) 
of Gd$_2$Ti$_2$O$_7$  may not be the previously suggested 4-${\bm k}$ structure \cite{stew}.
The results of a more in-depth numerical investigation of the low-temperature regime will be reported elsewhere.

{\it Discussion} --
Considering a  general symmetry-allowed anisotropic Hamiltonian,
we found that  ${\bm k}=(\frac{1}{2}\frac{1}{2}\frac{1}{2})$ partial
order can occur  over a wide range of anisotropic exchange and long-range dipolar
interactions in pyrochlore magnets. We  argued that fluctuations beyond s-MFT are responsible
for the stabilization of a 1-${\bm k}$ or 4-${\bm k}$ partially ordered structure.
This conclusion is based on results from E-TAP calculations where on-site fluctuations are included.
We used Monte Carlo simulations to illustrate that different values of the magnetic exchange
interactions can, as anticipated on the basis of the E-TAP calculations, lead to 
either 1-${\bm k}$ or 4-${\bm k}$ order.

From our work, we have exposed a likely mechanism  for the establishment of 
4-${\bm k}$ order in Gd$_2$Ti$_2$O$_7$ below its paramagnetic transition \cite{stew}.
Further quantitative progress on this problem will require a better estimate of the material exchange parameters 
\cite{Thompson_Yb2Ti2O7,balents,Savary_Er2Ti2O7}.
From this work we conclude that
the transition from the paramagnetic state to the 4-${\bm k}$ phase should be second order and belong to
the $3D$ $n=4$ cubic universality class. 
Experimental evidence \cite{Reotier_GTOcrit} suggests that this transition is second order in Gd$_2$Ti$_2$O$_7$.
Determining the critical exponents for this system could confirm our prediction but will  be a challenge, 
given that the exponents for the $3D$ $n=4$ cubic and $3D$ Ising, XY and Heisenberg universality 
classes are proximate to one another \cite{Pelissetto_review,carmona2000}.

Finally, the E-TAP method for frustrated magnets 
formulated here could prove useful for other systems where state 
selection {\it at}  $T_c$ is an open question. 
This is particularly so if the low-temperature state selected via (thermal or quantum) ObD 
is separated by a phase transition from the state selected at $T_c$.
In such a case,  an understanding of ObD at $T=0^+$  
can \emph{not} be leveraged to explain the selection at $T_c$.
Examples include the transition to long-range order in the 
pyrochlore Heisenberg antiferromagnet with
indirect DM interactions \cite{Elhajal2005,Canals_DM,Chern_DM},
the problem of  magnetization direction selection in face-centered cubic dipolar ferromagnets \cite{Roser_Corruccini,Bouchaud} and
the topical issue of state selection in $XY$ pyrochlore antiferromagnets 
\cite{Wong,McClarty_Er2Ti2O7,Savary_Er2Ti2O7,Champion_Er2Ti2O7,Wills_Er2Ti2O7,Stasiak_Er2Ti2O7,Zhitomirsky_Er2Ti2O7,Oitmaa_Er2Ti2O7}.


\begin{acknowledgements}

We thank Steve Bramwell, Alexandre Day, 
Jason Gardner, Paul McClarty, Oleg Petrenko,  Rajiv Singh, Ross Stewart,
 Ettore Vicari and Andrew Wills for useful discussions.
We acknowledge Pawel Stasiak for his help with the Monte Carlo simulations and 
Peter Holdsworth for his comments on the manuscript.
This work is supported by the NSERC of Canada,
the Canada Research Chair program (M.G., Tier 1) and by the Perimeter Institute for Theoretical Physics.
Research at PI is supported by the Government of Canada through Industry Canada and by the
Province of Ontario through the Ministry of Economic Development \& Innovation.

\end{acknowledgements}

\bibliography{ref}
\section*{\large SUPPLEMENTAL MATERIAL}
In this Supplementary Material we provide the reader with some of the technical details to assist with the reading of the main body of the paper. In Section I, we derive the Ginzburg-Landau theory presented in the main text. In Section II, we show that the two types of partially ordered states, 1-$\mathbfit k$  and 4-${\mathbfit k}$ states, have the same free energy in standard mean-field theory (s-MFT) with and without the presence of crystal electric field. Then, in Section III,  we show that the Ginzburg-Landau theory has an emergent $O(4)$ symmetry at the s-MFT level. In Section IV, we present the details of the extended TAP method (E-TAP) used in deriving the on-site fluctuation corrections needed to go beyond s-MFT.  In Section V, we go over the technical details of our Monte Carlo simulations. 
Finally, in Section VI, we consider the problem of $\{hhh\}$ degeneracy-lifting when the selection of
$\{\frac{1}{2}\frac{1}{2}\frac{1}{2}\}$ is induced by second nearest-neighbor exchange, ${\cal H}_2$,
 as opposed to long-range dipolar interaction (${\cal H}_{\rm dip})$. For
completeness sake, we consider a formulation of the spin Hamiltonian with
anisotropic exchange couplings as expressed in Ref.~[\onlinecite{balents, columbic,ETO, sungbin}].

\section{Derivation of the Ginzburg-Landau free energy}\label{GLT}

To proceed, we first recall the definition of the order parameters of the theory.
With the  spin direction at site $\bm{r}_i$ given by $\hat{e}_{\bm{k}_a}(\bm{r}_i)$ (Table \ref{tab:conf1k}), we 
introduce the order parameter $\psi_a$ for a particular ordering wavevector $\bm{k}_a$ as:
\begin{equation}
\label{eq2main}
\psi_{a}=  \frac{1}{N}\sum_{i}\bm{S}_i\cdot\hat{e}_{\bm{k}_{a}}(\bm{r}_i). 
\end{equation}

The Ginzburg-Landau free energy is constructed in terms of the four order parameters $\psi_a$ ($a=0,1,2,3$). 
Each 1-${\mathbfit k}$ order is a spin-density wave  and $\psi_a$ is the amplitude of the wave. 
The four order parameters $\psi_a$ form the components of a four-dimensional vector which fully describes 
the long-range order at the four $\{\frac{1}{2}\frac{1}{2}\frac{1}{2}\}$ momenta.  

To construct the free energy, we study how $\psi_a$ transforms under the space group symmetry of the pyrochlore lattice.
 Based on the real space spin configurations illustrated in Fig. 1 of the main text, we obtain how $\psi_a$ transforms. 
Under three fold rotations about the local $[111]$ direction of, say, the $0^{\rm th}$ sublattice (see Eq. (\ref{subvec})),
 three of $\psi_a$'s are permuted: $\psi_1\to \psi_2$, $\psi_2\to \psi_3$ and $\psi_3\to \psi_1$.
 Under a primitive FCC lattice translation, $\mathbfit r   \to\mathbfit r+\hat{x}/2+\hat{y}/2$ for example, two $\psi_a$'s, 
in this case $\psi_0$ and $\psi_3$, reverse sign. All $\psi_a$'s reverse sign under time-reversal transformation.

Lastly, the point inversion about a site on sublattice $a$ reverses the sign of $\psi_a$ while leaving other three order parameters intact. Since the symmetry operation is responsible for the elimination of an additional term in free energy invoked in Ref.~[\onlinecite{cepas2}], we demonstrate its consequence in detail here. 
Considering the 1-${\mathbfit k}$ structure with momentum $\mathbfit k_0$, 
the point inversion about a site on sublattice $0$ exchanges the spin configurations on adjacent kagome layers, 
which have opposite directions. As a result, $\psi_0 \to -\psi_0$. On the other hand, the point inversion about 
a site on any other sublattice than $0$ leaves the 1-${\mathbfit k}$ structure of $\mathbfit k_0$ intact.
 In other words, such a transformation leaves $\psi_0$ unchanged. Similar arguments can be made for $\mathbfit k_{a}$ with $a=1,2,3$. 

We now construct the Ginzburg-Landau free energy to quartic order. At the quadratic order, the only invariant term is $\sum_{a=0}^{3}\psi_a^2$. 
At quartic order, there are two terms that are invariant under the above symmetry transformations: 
$\left(\sum_{a=0}^3\psi_a^2\right)^2$ and $\sum_{a=0}^{3}\psi_a^4$. 
We note that the quartic term 
of the form
$\sim u_3 \psi_0\psi_1\psi_2\psi_3$ invoked
in Ref.~[\onlinecite{cepas2}] 
is odd under the 
aforementioned
point inversion symmetry and this term should not be present in the free energy. Given these considerations, we write down the Ginzburg-Landau free energy:
\begin{equation}
\label{eq:landau}
{\cal {F}}=\frac{1}{2}r\sum_{a=0}^{3}\psi_a^2+u\left(\sum_{a=0}^3\psi_a^2\right)^2+v\sum_{a=0}^{3}\psi_a^4. 
\end{equation}
In this work, 
we make use of the following quantities:
\begin{subequations}
\begin{eqnarray}\label{opp}
\bm{\psi}		&\equiv&	(\psi_0,\psi_1,\psi_2,\psi_3) ,	\\ \label{opp11}
|\bm{\psi}|		&=&		\sqrt{\sum_{a=0}^3\psi_a^2} ,		\\ \label{opp12}
\hat{{\psi}}	&\equiv&	\frac{\bm{\psi}}{|\bm{\psi}|} . 	\label{opp13}
\end{eqnarray}
\end{subequations}
Here, $\bm {\psi}$ is the 4-component vector order paramete and
$\vert \bm {\psi} \vert$ is its magnitude.
$\hat{{\psi}}$ is a 4-component unit vector parallel to  $\bm {\psi}$ which is convenient to parametrize the behavior of the system 
below the critical temperature, $T_c$, in the Monte Carlo simulations discussed in Section V below.


\section{Free-energy degeneracy of 1-${\mathbfit k}$ and 4-${\mathbfit k}$ At the s-MFT Level}

In this section we show that, at the s-MFT level, the 1-${\mathbfit k}$ and 4-${\mathbfit k}$ states
 have the same free energy with and without considering the effect of crystal electric field in pyrochlore lattice.

\subsection{No Crystal Electric Field}\label{NOCEF}
We begin by defining the orientation of the  moments, $\hat{e}^{\alpha}_{\mathbfit{k}_a}$, 
on a typical tetrahedron that makes up a 1-$\mathbfit k$ state. Here, $\alpha=0,\cdots,3$ 
corresponds to one of the four bases (sublattices) of the pyrochlore structure 
(see Eq. (\ref{subvec})) and $\bm k_a$ refers to ordering wave vector of the corresponding 1-$\bm k$ state. 
 $\hat{e}^{\alpha}_{\mathbfit{k}_a}$'s are given in Table \ref{tab:conf1k} and are expressed in global Cartesian coordinates.

\begin{widetext}
\begin{center}
\begin{table}
    \begin{tabular}{ | l | l | l | l | l |}
    \hline
    ${\mathbfit k_a}$ & sublattice 0 & sublattice 1  & sublattice 2  & sublattice 3 \\ \hline
    $ \mathbfit k_0=(\frac{1}{2},\frac{1}{2},\frac{1}{2}) $ &$\hat {e}_{\mathbfit k_0}^0=(0, 0, 0)$ & $\hat {e}_{\mathbfit k_0}^1=(\frac{-1}{\sqrt{2}},  \frac{1}{\sqrt{2}}, 0)$ &  $\hat {e}_{\mathbfit k_0}^2=(\frac{1}{\sqrt{2}},  0,  \frac{-1}{\sqrt{2}})$ & $\hat {e}_{\mathbfit k_0}^3=(0, \frac{-1}{\sqrt{2}},  \frac{1}{\sqrt{2}})$ \\ \hline
    $\mathbfit k_1=(-\frac{1}{2},-\frac{1}{2},\frac{1}{2} )$&$ \hat{ e}_{\mathbfit k_1}^0=(\frac{1}{\sqrt{2}},  \frac{-1}{\sqrt{2}}, 0)$ & $\hat{ e}_{\mathbfit k_1}^1=(0, 0, 0)$ & $\hat{ e}_{\mathbfit k_1}^2=(0, \frac{1}{\sqrt{2}},  \frac{1}{\sqrt{2}})$ & $\hat{ e}_{\mathbfit k_1}^3=(\frac{-1}{\sqrt{2}},  0,  \frac{-1}{\sqrt{2}})$\\ \hline
    $\mathbfit k_2=(-\frac{1}{2},\frac{1}{2},-\frac{1}{2})$ & $\hat{ e}_{\mathbfit{k_2}}^0=(\frac{-1}{\sqrt{2}},  0,  \frac{1}{\sqrt{2}})$ &$\hat{ e}_{\mathbfit{k_2}}^1=(0, \frac{-1}{\sqrt{2}},  \frac{-1}{\sqrt{2}})$ &$\hat{ e}_{\mathbfit{k_2}}^2=(0, 0, 0)$ & $ \hat{ e}_{\mathbfit{k_2}}^3=(\frac{1}{\sqrt{2}},  \frac{1}{\sqrt{2}}, 0)$ \\ \hline
    $ \mathbfit k_3=(\frac{1}{2},-\frac{1}{2},-\frac{1}{2})$ & $\hat{ e}_{\mathbfit k_3}^0=(0, \frac{1}{\sqrt{2}},  \frac{-1}{\sqrt{2}})$ &$ \hat{ e}_{\mathbfit k_3}^1=(\frac{1}{\sqrt{2}},  0,  \frac{1}{\sqrt{2}})$ & $\hat{ e}_{\mathbfit k_3}^2=(\frac{-1}{\sqrt{2}},  \frac{-1}{\sqrt{2}}, 0)$ & $\hat{ e}_{\mathbfit k_3}^3=(0, 0, 0)$ \\ \hline
    \end{tabular}
   \caption{Spin configurations of a single tetrahedron in 1-${\mathbfit k}$ states}
  \label{tab:conf1k}
 \end{table}  
\end{center}
\end{widetext}

The orientation of the rest of the moments in a 1-${\mathbfit k}$ state can be obtained from a single tetrahedron configuration using the following equation:
\begin{equation}
\label{eq:dir}
\hat {e}_{\mathbfit k_a}(\mathbfit r_i) = \hat{ e}_{\mathbfit k_a}^{\alpha} \cos(\mathbfit k_a \cdot\mathbfit R_i)
\end{equation}
where  $\mathbfit R_i\equiv \mathbfit r_i-\mathbfit r^{\alpha}$ are the FCC lattice position vectors, 
$\mathbfit r_i$ are pyrochlore lattice position vector.  
The vector $\mathbfit r^{\alpha}$ specifies the bases' positions in pyrochlore lattice and are given by:
\begin{subequations}\label{subvec}
\begin{eqnarray}
\mathbfit r^0&=&( 0, 0, 0)\\
\mathbfit r^1&=&( \frac{1}{\sqrt{2}},  \frac{1}{\sqrt{2}}, 0)\\
\mathbfit r^2&=&(  \frac{1}{\sqrt{2}}, 0,  \frac{1}{\sqrt{2}})\\
\mathbfit r^3&=&( 0,  \frac{1}{\sqrt{2}},  \frac{1}{\sqrt{2}}).
\end{eqnarray}
\end{subequations}
The above coordinates are expressed in units of $r_{\mathrm{nn}}$, the nearest-neighbor distance in pyrochlore structure.
The spin orientations corresponding to the
4-${\mathbfit k}$ states can be expressed in terms of linear combinations of the four 1-${\mathbfit k}$ states:
\begin{equation}
\label{lincomb}
\hat {e}_{4k}(\mathbfit r_i)\equiv\sum_{a=0}^3 \hat{\psi}_a \hat{e}_{\mathbfit k_a}(\mathbfit r_i). 
\end{equation} 
Here $ \hat{\psi}_a=\pm1/2$ conveniently keeps $\hat {e}_{4k}(\mathbfit r_i)$ normalized (except for the sites with zero moments) and generate all possible 4-$\mathbfit k$ states.

Having defined the 1-$\bm k$ and 4-$\bm k$ states, we now proceed to show that both the 1-${\mathbfit k}$ and 4-${\mathbfit k}$ 
states have the same s-MFT free energy. The general form of the s-MFT free energy  for a classical 3-component spin, ${\bm S}_i$ at site $i$ reads\cite{Matt1}:
\begin{eqnarray}
\label{eq:fenergy}
F& = & -\frac{1}{2}\sum_{i, j}m^{\mu}_i J_{ij}^{\mu\nu}m^{\nu}_j-\frac{1}{\beta}\sum_i\ln\big(4\pi\frac{\sinh(\beta h_i)}{\beta h_i}\big). \nonumber \\
&&
\end{eqnarray}
$J_{ij}^{\mu\nu}$ is the bilinear spin-spin exchange coupling matrix defined through the Hamiltonian of Eq. (1) in the main text. 
The magnetic moment, $\mathbfit m_i$ for both 1-$\mathbfit k$ and 4-$\mathbfit k$ states at site $\bm r_i$ ($i$) 
is obtained from the s-MFT self-consistent (Langevin function) equation, 
 \begin{equation}
 \label{selfi}
 \mathbfit m_i =- \frac{\mathbfit h_i}{h_i}[\coth(\beta h_i)-\frac{1}{\beta h_i}]
 \end{equation}
 where
 \begin{equation}
 \label{magloc}
 h_i\equiv|\mathbfit h_i|=|\sum_{j,\nu}J_{ij}m_j^{\nu}| .
 \end{equation}

In what follows, we compare the free energy, $F$, within standard mean-field theory (s-MFT), 
of a 1-$\bm k$ state, with 4-component order parameter, 
${\bm \psi} = (|{\bm {\psi}}|, 0, 0, 0)$,
with the free energy of a 4-$\bm k$ state, with 4-component order parameter
 ${\bm \psi} =(|{\bm {\psi}}|/2, |{\bm {\psi}}|/2, |{\bm {\psi}}|/2, |{\bm {\psi}}|/2)$.
We thus have $|\mathbfit m_i|=|\bm{\psi}|$ which is also confirmed numerically by solving the Eq. (\ref{selfi}). We define 
for a 1-$\mathbfit k$ state:
\begin{equation}
\label{mag-1k} 
\mathbfit m_i=|\bm{\psi}|\hat {e}_{\mathbfit k_a}(\mathbfit r_i)\equiv\mathbfit m_{\mathbfit k_a}(\mathbfit r_i)
 \end{equation} 
while we have for a 4-$\mathbfit k$ state:
 \begin{equation}
 \label{mag-4k}
\mathbfit m_i=|\bm{\psi}|\hat{e}_{4k}(\mathbfit r_i)=\sum_{a=1}^4 \hat{\psi}_a \mathbfit m_{\mathbfit k_a}(\mathbfit r_i)\equiv \mathbfit m_{4k}(\mathbfit r_i).
 \end{equation} 
 In Eq. (\ref{eq:fenergy}), we are employing implicit summation convention for repeated Greek superscripts $\mu$ and $\nu$ which represent Cartesian coordinates.  
 
We now show that both terms in Eq. (\ref{eq:fenergy}) are the same for 1-$\bm k$ and 4-$\bm k$ states and thus, at a given temperature
the free-energy of 1-${\bm k}$ and 4-${\bm k}$ is the same at the s-MFT level.

\vspace{2mm}

{\bf First Term of Eq. (\ref{eq:fenergy}):}  $--$ 
The first term,  $F_1\equiv -\frac{1}{2}\sum_{i, j}m^{\mu}_i J_{ij}^{\mu\nu}m^{\nu}_j$, reads for a typical 1-${\mathbfit k}$ state:
\begin{equation}
\label{f1k-r}
F_1(1k)\equiv - \frac{1}{2}\sum_{ij}m^{\mu}_{\mathbfit k_a}(\mathbfit r_i)J^{\mu\nu}_{ij}m^{\nu}_{\mathbfit k_a}(\mathbfit r_j)
\end{equation}
Using Eqs. (\ref{eq:dir}, \ref{mag-1k}) we have:
\begin{equation}
\label{f1k-q}
F_1(1k)=-\frac{1}{2}\sum_{\alpha\beta}m_{\mathbfit k_a}^{\mu}(\mathbfit r^{\alpha})J^{\mu\nu}_{\alpha\beta}(\mathbfit k_a)m_{\mathbfit k_a}^{\nu}(\mathbfit r^{\beta})
\end{equation}
where $\alpha$ and $\beta$ are sublattice labels and $J^{\mu\nu}_{\alpha\beta}(\mathbfit k_a)$ is the Fourier transform of $J^{\mu\nu}_{ij}$. Similarly, using Eqs. (\ref{mag-4k}, \ref{f1k-r}),  $F_1$ for a 4-${\mathbfit k}$ state can be written as:
\begin{equation}
F_1(4k)=-\frac{1}{2}\sum_{ij}\sum_{ab}\hat{\psi}_a\hat{\psi}_bm^{\mu}_{\mathbfit k_a}(\mathbfit r_i)J_{ij}^{\mu\nu}m^{\nu}_{\mathbfit k_b}(\mathbfit r_j).
\end{equation}
Using Eq. (\ref{eq:dir}) we obtain:
\begin{equation}
\label{E1kE4k}
F_1(4k)=-\frac{1}{2}\sum_{ab;\alpha\beta}\hat{\psi}_a\hat{\psi}_b m^{\mu}_{\mathbfit k_a}(\mathbfit r^{\alpha})J^{\mu\nu}_{\alpha\beta}(\mathbfit k_a)m^{\nu}_{\mathbfit k_b}(\mathbfit r^{\beta})\delta_{ab} .
\end{equation}
Consequently,
\begin{equation}
F_1(4k)=-\frac{1}{2}\sum_{a;\alpha\beta}(\hat{\psi}_a)^2 m_{\mathbfit k_a}^{\mu}(\mathbfit r^{\alpha})J^{\mu\nu}_{\alpha\beta}(\mathbfit k_a)m_{\mathbfit k_a}^{\nu}(\mathbfit r^{\beta})
\end{equation}
and using Eq. (\ref{f1k-q}), we find:
\begin{eqnarray}
F_1(4k)&=&\sum_a(\hat{\psi}_a)^2F_1(1k).
\end{eqnarray}
Considering that for all $a$,  $\hat{\psi}_a=\pm\frac{1}{2}$, and all $F_1(1k)$ are the same, we thus have:
\begin{eqnarray}
\label{e1k4k}
F_1(4k)&=&F_1(1k).
\end{eqnarray}

{\bf Second Term of Eq. (\ref{eq:fenergy}):} $--$ 
In this term, the only variable that depends on different spin configurations is the magnitude of local field, $h_i$, 
in Eq. (\ref{magloc}),which we focus on. Using Eq. (\ref{mag-1k}), the local field, $\mathbfit h_i$, experienced by a moment at site $\mathbfit r_i$ in a 1-$\mathbfit k$ state,
is given by:
\begin{equation}\label{lf1k-2}
 h^{\mu}_{\mathbfit k_a}(\mathbfit r_i)\equiv\sum_{j\nu}J_{ij}^{\mu\nu}m^{\nu}_{\mathbfit k_a}(\mathbfit r_j),
\end{equation}
while for a 4-$\mathbfit k$ state we have
\begin{equation}
\label{loclin}
 h^{\mu}_{4k}(\mathbfit r_i)\equiv\sum_{j\nu}J_{ij}^{\mu\nu}m^{\nu}_{4k}(\mathbfit r_j)=\sum_a\hat{\psi}_ah^{\mu}_{\mathbfit k_a}(\mathbfit r_i).
\end{equation}
where $m^{\nu}_{4k}(\mathbfit r_j)=\sum_a\hat{\psi}_a m^{\nu}_{\mathbfit k_a}(\mathbfit r_j)$.  
Within s-MFT, the local field at each site is antiparallel to the moment at that site. As a result, $\mathbfit h_{\mathbfit k_a}(\mathbfit r_i)$ can be written as
\begin{eqnarray}
\label{lf1k-1}
\mathbfit h_{\mathbfit k_a}(\mathbfit r_i)&=&C\mathbfit m_{\mathbfit k_a}(\mathbfit r_i).
\end{eqnarray}
 Here, $C$ is the same constant for each of the 1-$\mathbfit k$ states with ${\bm k} \equiv {\bm k}_a$ ($a=0,1,2,3$),
since all of 1-$\mathbfit k$ states have the same free energy by definition.

Similarly, for the 4-$\mathbfit k$ state, using Eqs. (\ref{loclin}, \ref{lf1k-1}), we find
\begin{equation}\label{lf4k}
\mathbfit h_{4k}(\mathbfit r_i)=C\sum_{a} {\hat {\psi_a}} \mathbfit m_{\mathbfit k_a}(\mathbfit r_i)=C\mathbfit m_{4k}(\mathbfit r_i) .
\end{equation}
Since we have $|\bm\psi|$ the same for both 1-$\mathbfit k$ and 4-$\mathbfit k$ states by construct, 
Eq. (\ref{lf4k}) shows that the magnitude of the local field is the same for 1-${\mathbfit k}$ and 4-${\mathbfit k}$ at the lattice sites with nonzero moments
at a given temperature $T$. We note that, the theory is internally consistent, since we have taken 
$|\bm\psi|$ to be the same for both 1-$\mathbfit k$ and 4-$\mathbfit k$ states which is confirmed numerically as mentioned earlier.
As a result, the second term in Eq. (\ref{eq:fenergy}) is the same for 1-${\mathbfit k}$ and 4-${\mathbfit k}$. 

Based on Eqs. (\ref{e1k4k}, \ref{lf4k}) and Eq. (\ref{eq:fenergy}), we therefore find that 
the s-MFT free energy is the same for the 1-$\mathbfit k$ and 4-$\mathbfit k$ states.

\subsection{With Crystal Electric Field}

The discussion in the previous subsection, as well as in the main text, has assumed for simplicity that the 
spin degrees of freedom, ${\bm S}_i$ at each site $i$, is a classical vector of fixed length. As a result, the s-MFT treatment leads for
the self-consistent equation for ${\bm m}_i$ to the Langevin function in Eq. (\ref{selfi}). Had we assumed a quantum ${\bm S}_i$, Eq. (\ref{selfi})
 would be replaced $\mathbfit m_i =- \frac{\mathbfit h_i}{h_i}B_{S}(h_j)$, where $B_{S}(x) = \frac{2S+1}{2S}\coth(\frac{2S+1}{2S}x) - \frac{1}{2}\coth(\frac{x}{2S})$
is the Brillouin function.
In this case, one can repeat the argument of the previous subsection and, again, show that 
the 1-${\bm k}$ and 4-${\bm k}$ states are degenerate.

One may then ask whether the single-ion anisotropy, arising from the crystal electrical field (CEF) 
effect, may change the conclusion that the 1-${\bm k}$ and 4-${\bm k}$ states are degenerate.
Again, such a question can be asked for either classical or quantum spins ${\bm S}_i$.
We now proceed to briefly show that introducing a crystal field anisotropy does not lift the
degeneracy between the  1-${\bm k}$ and 4-${\bm k}$ states.

At the s-MFT level, the Hamiltonian including crystal electric field or single ion anisotropy ($H_{\mathrm{CEF}}$) can be written as:

\begin{equation}
\label{crysHamil}
H_{\mathrm{MF}}=\sum_{i=1}^N\Big[\mathbfit h_i \cdot \mathbfit S_i +  H_{\mathrm{CEF}}(\mathbfit S_i) -\frac{1}{2}\bm h_i\cdot\bm m_i\Big],
\end{equation}
where $H_{\mathrm{CEF}}$ can be expressed in terms of Stevens' operators according to the symmetry of pyrochlore structure\cite{Gardner_RMP}:

\begin{equation}
\label{crys}
H_{\mathrm{CEF}}( \mathbfit S_i)= \sum_{n=2,4,6}\sum_{m=0}^6B_n^mO_n^m(\mathbfit S_i).
\end{equation}
We consider $\mathbfit S_i$ expressed in terms of its components in the local [111]  coordinate system. 

The s-MFT free energy can be written as
\begin{eqnarray}
\label{crysfree}
F&=&-\frac{1}{\beta}\sum_{i=1}^N\ln(Z_i),
\end{eqnarray}
where
\begin{equation}
\label{cryspart}
Z_i= \mathrm{Tr}[\exp\big(-\beta\mathbfit h_i \cdot {\mathbfit S_i} -\beta H_{\mathrm{CEF}}({\mathbfit S_i})+\frac{\beta}{2}\bm h_i\cdot\bm m_i \big)].
\end{equation}
In what follows, we aim to compare the s-MFT free energy of 1-$\bm k$ and 4-$\bm k$ states using Eq. (\ref{crysfree}) without explicitly finding its global minimum. 
In Eq. (\ref{cryspart}), $\bm S_i$ can be treated as a classical vector or a quantum mechanical operator. However, we show that the outcome is independent of this choice. The result of performing the $\mathrm{Tr}$ in Eq. (\ref{cryspart}) can be written in general form of 
\begin{equation}
\label{form1}
\exp(\frac{\beta}{2}\bm h_i\cdot\bm m_i )\mathcal{G}(h_i^x, h_i^y, h_i^z, \beta).
\end{equation}
Since we are comparing the free energy of the 1-$\mathbfit k$ and 4-$\mathbfit k$ states, 
$\mathbfit h_i$ and $\bm m_i$ are known for every $i$. So, we have
 
 \begin{equation}
 \exp(\frac{\beta}{2}\bm h_i\cdot\bm m_i )=\exp(\frac{\beta C|\bm{\psi}|^2}{2})\equiv \tilde{C},
 \end{equation}
which is the same for both 1-$\bm k$ and 4-$\bm k$ states. Here, $C$ is the same constant in Eqs. (\ref{lf1k-1}, \ref{lf4k}). As a result, Eq. (\ref{form1}) can be rewritten as
\begin{equation}
\tilde{C}\tilde{\mathcal{G}}(h_i, \theta_i, \phi_i), 
\end{equation}
 where $\theta_i$ and $\phi_i$ are the polar and azimuthal angles of  $\mathbfit h_i$ in the local [111] frame of coordinates. We dropped $\beta$ from the argument of $\tilde{\mathcal{G}}$ for simplicity.
 Now we show the $\tilde{\mathcal{G}}$ is the same for both 1-$\mathbfit k$ and 4-$\mathbfit k$ states. According to Eq. (\ref{lf4k}),  $h_i$ is the same for both 1-$\mathbfit k$ and 4-$\mathbfit k$ states. Considering Table \ref{tab:conf1k} and  Eq. (\ref{lincomb}), $\forall i$,  $\theta_i=\pi/2$ for both of these states. On the other hand,  $\phi_i$ can only have the following values: $\{\frac{\pi}{2}, \frac{3\pi}{2}, \frac{\pi}{6}, \frac{5\pi}{6}, \frac{7\pi}{6}, \frac{11\pi}{6}\}$ for both 1-$\bm k$ and 4-$\bm k$ states.  Although for 1-$\mathbfit k$ and 4-$\mathbfit k$ states, these angles are distributed differently over the lattice, all of them are present in a typical 1-$\bm k$ and 4-$\bm k$ states. Since there is a sum over all lattice sites in Eq. (\ref{crysfree}), the outcome of this equation is the same for the 1-$\mathbfit k$ and 4-$\mathbfit k$ states. Consequently, including the crystal electric field in 
the s-MFT Hamiltonian does not differentiate between 1-$\bm k$ and 4-$\bm k$ states, free-energy-wise.

We numerically confirmed this result by minimizing Eq. (\ref{crysfree}) and solving the 
resulting s-MFT self-consistent equation where $\bm S_i$ were spin-$7/2$ quantum mechanical operators and $B_n^m$ coefficients in Eq. (\ref{crys}), were chosen in accordance with the experimental values for Gd$_2$Ti$_2$O$_7$ from Ref. [\onlinecite{Glazkov_CEF}]. We found that the 1-${\bm k}$ and 4-${\bm k}$ states have identical paramagnetic critical transition temperature,$T_c$, and identical free energy below $T_c$.

\section{Emergent $O(4)$ symmetry up to quartic order in s-MFT}

In this section we demonstrate that the free energy within the s-MFT treatment is $O(4)$ symmetric up to quartic order in the $\psi_a$'s. To do this, we expand the s-MFT free energy, Eq. (\ref{eq:fenergy}), in powers of $h_i$:
\begin{equation}
\label{hpansion}
F=\sum_{n=0}^{\infty}\sum_{i}\frac{1}{n!}\frac{\partial^{n}F}{\partial h_i^n}\Big|_{h_i=0} h_i^n
\end{equation}
where $h_i$ is defined in Eq. (\ref{magloc}). Considering the pyrochlore structure, $\sum_i$ is equivalent to $\sum_{\alpha=0}^3\sum_{{\mathbfit R}_i}$, where $\alpha$ is the sublattice label and $\mathbfit R_i$ is the FCC lattice position vector.  As a result, $h_i$ can be relabelled as $h^{\alpha}(\mathbfit R_i)$. This notation makes the book-keeping clearer 
in what follows. We will focus on the $n=4$ term because, based on our Ginzburg-Landau symmetry analysis in Section \ref{GLT}, the quadratic term is $O(4)$ invariant.

For $T \lesssim T_c$ and in a region of parameter space with $\{\frac{1}{2}\frac{1}{2}\frac{1}{2}\}$ ordering wave vectors, $\mathbfit h^{\alpha}(\mathbfit R_i)$ can be written as
\begin{equation}
\label{linfcomb}
\mathbfit h^{\alpha}(\mathbfit R_i) = \sum_a\hat{\psi}_a\mathbfit h_{\mathbfit k_a}(\mathbfit r_i),
\end{equation}
where $\bm{r}_i=\bm{R}_i+\bm{r}^{\alpha}$. Using Eqs. (\ref{eq:dir}, \ref{mag-1k}, \ref{lf1k-2}) we have
 \begin{equation}
\label{eq:dir1}
\mathbfit h_{\mathbfit k_a}(\mathbfit r_i)= |\mathbfit h_{\mathbfit k_a}(\mathbfit r_i)|\hat{ e}_{\mathbfit k_a}^{\alpha} \cos(\mathbfit k_a \cdot \mathbfit R_i),
\end{equation}
where according to the arguments presented in Section \ref{NOCEF}, $|\mathbfit h_{\mathbfit k_a}(\mathbfit r_i)|$ is the same for all sites with nonzero moments.  So  $|\mathbfit h_{\mathbfit k_a}(\mathbfit r_i)|\equiv h$. The fourth order term in $h_i$ arising in Eq. (\ref{hpansion}), 
can be written as $\propto \sum_{\mathbfit R_i}\sum_{\alpha=0}^3|\mathbfit h^{\alpha}(\mathbfit R_i)|^4$. 
We now proceed to show that this term is $O(4)$ invariant.


Using Eqs. (\ref{linfcomb}, \ref{eq:dir1}), we have:
\begin{widetext}
\begin{equation}
\label{maiproof1}
\sum_{\mathbfit R_i}\sum_{\alpha=0}^3|\mathbfit h^{\alpha}(\mathbfit R_i)|^4=\sum_{\mathbfit R_i}\sum_{\alpha}h^4\left(\sum_{b=0}^3\hat{\psi}_b^2+\sum_{b_1<b_2} 2\hat{\psi}_{b_1}\hat{\psi}_{b_2} \left(\hat{ e}_{\mathbfit k_{b_1}}^{\alpha}\cdot\hat{ e}_{\mathbfit k_{b_2}}^{\alpha}\right)\cos(\mathbfit k_{b_1} \cdot \mathbfit R_i)\cos(\mathbfit k_{b_2} \cdot \mathbfit R_i)\right)^2 
\end{equation}
\begin{equation}
\label{maiproof2}
\sum_{\mathbfit R_i}\sum_{\alpha=0}^3|\mathbfit h^{\alpha}(\mathbfit R_i)|^4 =\frac{Nh^4}{4}\sum_{\alpha=0}^3[(\sum_{b=0}^3\hat{\psi}_b^2)^2+\sum_{b_{1}<b_{2}}\hat{\psi}_{b_1}^2\hat{\psi}_{b_2}^2],
 \end{equation}
 \begin{equation}
 \label{maiproof3}
\sum_{\mathbfit R_i}\sum_{\alpha=0}^3|\mathbfit h^{\alpha}(\mathbfit R_i)|^4=\frac{3Nh^4}{4|\bm{\psi}|^4}(\sum_{b=0}^3\psi_b^2)^2. 
\end{equation}
\end{widetext}

In Eq. (\ref{maiproof1}), we used the relation that $\hat{e}^{\alpha}_{\mathbfit k_{b_1}}\cdot \hat{e}^{\alpha}_{\mathbfit k_{b_2}}=-1/2$ for $b_1\neq b_2\neq \alpha$ (see Table \ref{tab:conf1k}).
 We use momentum conservation going from Eq. (\ref{maiproof1}) to Eq. (\ref{maiproof2}). In Eq. (\ref{maiproof3}), we  used the relation $\psi_b=|\bm{\psi}|\hat{\psi}_b$.
$N$ is the number of sites. 

\section{Extended TAP method}

The extented TAP method (E-TAP) developed in this work provides a systematic methodology 
to compute the corrections beyond s-MFT to all orders of inverse temperature, $\beta$.
Our approach builds on the method described by Georges and Yedidia \cite{Yadidah}. 
However, our focus is on the temperature regime close to the transition, $T_c$. 
In the rest of this section, 
we derive the E-TAP equations for Heisenberg spins of a general bilinear spin Hamiltonian. 

We consider the Gibbs free energy:
\begin{equation}
\label{Gibbs1}
G=\frac{-\ln(\text{Tr}[\exp\big(-\beta {\mathcal{H}}+ \sum_i \bm \lambda_i \cdot (\mathbfit S_i-\mathbfit m_i)\big)])}{\beta}
\end{equation}
with
\begin{equation}
\label{Ham1}
{\mathcal{H}}=\frac{1}{2}\sum_{ij}S_i^{\mu}J_{ij}^{\mu\nu} S_j^{\nu},
\end{equation}
where $\mathbfit m_i$ is the magnetization at site $i$, $\mathbfit m_i\equiv\langle \mathbfit S_i \rangle$,  and $\bm \lambda_i$ is a Lagrange multiplier which enforces $\mathbfit m_i$ to remain at its mean-field expectation value ($\mathbfit m_i=\langle \mathbfit S_i \rangle$) \cite{Yadidah}. Note that a factor of $\beta$ is implicit in $\bm \lambda_i$ i.e. $\bm \lambda_i = \beta\bm h_i$,
where $\bm h_i$ is defined in Eq. (\ref{magloc}). ${\mathcal{H}}$ is the Hamiltonian considered in Eq. (1) of the main text. In Eq. (\ref{Ham1}), Greek labels represent Cartesian coordinates and implicit summation convention is again employed.

A Taylor series expansion of Eq. (\ref{Gibbs1}) in powers of $\beta$ reads:
\begin{eqnarray}
\label{expansion1}
 G(\beta)&=& \frac{1}{\beta}\Big(\tilde{G}(\beta)\big|_{\beta=0} +\frac{\partial \tilde{G}(\beta)}{\partial\beta}\Big|_{\beta=0}\beta \\ \nonumber
 &&+\frac{1}{2!}\frac{\partial^2 \tilde{G}(\beta)}{\partial\beta^2}\Big|_{\beta=0}\beta^2\cdots\Big).
 \end{eqnarray}
 where $\tilde{G}(\beta)\equiv\beta G(\beta)$.
We define,
\begin{eqnarray}
U&\equiv&\frac{1}{2}\sum_{ij}\delta S_i^{\mu}J_{ij}^{\mu\nu}\delta S_j^{\nu} 
\end{eqnarray}
where
\begin{eqnarray}
\delta S_j^{\nu}&\equiv&S_j^{\nu}- m_j^{\nu}.
\end{eqnarray}
It can be shown that the derivatives of $\tilde{G}(\beta)$ with respect to $\beta$ can be evaluated in terms of expectation value of powers of $U$\cite{Yadidah}. We find:
\begin{subequations}
\begin{eqnarray}
\frac{\partial\tilde{G}(\beta)}{\partial\beta}&=&\langle H\rangle \label{TAPS1} \\
\frac{\partial^2 \tilde{G}(\beta)}{\partial\beta^2}&=& -\langle U^2\rangle \label{TAPS2} \\
\frac{\partial^3\tilde{G}(\beta)}{\partial\beta^3}&=& \langle U^3 \rangle \label{TAPS3}
\end{eqnarray}
\end{subequations}
The $\langle \cdots \rangle$ above denotes a thermal average which, for a general observable $O$, is given by:
\begin{equation}
\label{overage}
\langle O\rangle=\frac{\text{Tr}[O \exp\big(-\beta H+ \sum_i \bm \lambda_i \cdot (\mathbfit S_i-\mathbfit m_i)\big) ]}{\text{Tr}[\exp\big(-\beta H+ \sum_i \bm \lambda_i \cdot (\mathbfit S_i-\mathbfit m_i)\big)]}.
\end{equation}
According to Eq. (\ref{expansion1}), all the averages are calculated at $\beta=0$. The limit $\beta=0$ in Eq. (\ref{overage}) corresponds to the s-MFT approximation and,
consquently, $\langle \cdots \rangle$ is changed to $\langle \cdots \rangle_{\mathrm{MF}}$. The first two terms in Eq. (\ref{expansion1}) correspond to the s-MFT free energy while the higher order terms in $\beta$ provide the corrections beyond s-MFT and introduce the fluctuations.
 
 Calculating the expectation value of powers of $U$ reduces to the evaluation of averages of the following form:
\begin{equation}
\label{average1}
 \langle \delta S^{\alpha_1}_{i_1}\delta S^{\alpha_2}_{i_2}\cdots\delta S^{\alpha_n}_{i_n}\rangle_{\mathrm {MF}}
 \end{equation}
 where $i_n$ represents the site label, $\alpha_n$ represents a Cartesian coordinate and $n$ is the number of $\delta S$'s in the Eq. (\ref{average1}). We henceforth drop the $\mathrm {MF}$ subscript for simplicity. 
 For $n=1$, the expectation value in Eq. (\ref{average1}) is zero due to the relation  $\mathbfit m_i=\langle \mathbfit S_i \rangle$. For $ n \geq 2 $, however, this expectation value is nonzero only if no site label appears only once.
For example, averages of the following form have a nonzero contribution:
 \begin{equation}
 \langle \delta S^{\alpha_1}_{i}\delta S^{\alpha_2}_{i}\delta S^{\alpha_3}_{j}\delta S^{\alpha_4}_{j}\delta S^{\alpha_5}_{j}\rangle= \langle \delta S^{\alpha_1}_{i}\delta S^{\alpha_2}_{i}\rangle\langle\delta S^{\alpha_3}_{j}\delta S^{\alpha_4}_{j}\delta S^{\alpha_5}_{j}\rangle
 \end{equation}
 
  The expectation values above can be calculated using the self-consistent s-MFT equation which, for the case of 3-component classical spins, is given by,
  \begin{equation}
 \label{slfMF1}
\mathbfit m_i = -\frac{\bm \lambda_i}{|\bm \lambda_i|}[\coth(|\bm \lambda_i|)-\frac{1}{|\bm \lambda_i|}].
\end{equation}
Consequently 
 \begin{equation}
 \label{chi1}
 \langle\delta S^{\alpha}_i\delta S^{\beta}_i\rangle = \frac{\partial m^{\alpha}_i}{\partial \lambda^{\beta}_i} = \chi_i^{\alpha\beta} ,
 \end{equation}
 \begin{equation}
 \langle\delta S^{\alpha}_i\delta S^{\beta}_i\delta S^{\gamma}_i\rangle = \frac{\partial\chi_i^{\alpha\beta}}{\partial \lambda^{\gamma}_i},
 \end{equation}
and, generally,
\begin{equation}
 \langle \delta S^{\alpha_1}_{i}\delta S^{\alpha_2}_{i}\cdots\delta S^{\alpha_n}_{i}\rangle=\frac{\partial \langle \delta S^{\alpha_1}_{i}\delta S^{\alpha_2}_{i}\cdots\delta S^{\alpha_{n-1}}_{i}\rangle}{\partial \lambda_i^{\alpha_n}}.
 \end{equation} 
 
Referring to Eq. (\ref{expansion1}), the lowest order correction beyond s-MFT 
originates from the third term. 
Considering Eq. (\ref{TAPS2}), the result reads:
 \begin{equation}
 \label{TAP2}
\Omega\equiv\frac{\beta}{2!}\frac{\partial^2\tilde{G}}{\partial \beta^2}\Big|_{\beta=0}=-\frac{\beta}{4}\sum_{ij} \sum_{\alpha\beta\gamma\delta} J^{\alpha\gamma}_{ij} J^{\beta\delta}_{ij} \chi^{\alpha\beta}_i  \chi^{\gamma\delta}_j.
\end{equation}
In our expansion, we are interested in a temperature range close to $T_c$. In this case,  Eq. (\ref{slfMF1}) can be expanded for small $|\bm \lambda_i|$:
\begin{equation}
\label{eq1}
m^{\alpha}_i=-\frac{\lambda^{\alpha}_i}{3}[1- \frac{(|\bm \lambda_i|)^2}{15} + \frac{2(|\bm \lambda_i|)^4}{315} + \cdots ],
\end{equation}
To calculate $\chi^{\alpha\beta}_i$ in Eq. (\ref{chi1}), we differentiate Eq. (\ref{eq1}) with respect to $\lambda^{\alpha}_i$. This gives:
\begin{eqnarray}
\label{chiexpan}
\chi^{\alpha\beta}_i &= &-\frac{\delta_{\alpha\beta}}{3}(1- \frac{|\bm \lambda_i|^2}{15} )
+ \frac{2}{25}\frac{\lambda^{\alpha}_i\lambda^{\beta}_i}{3} +  \cdots,
\end{eqnarray}
where $\cdots$ means higher order terms. 
%
%
According to the general discussion pertaining to the Ginzburg-Landau theory presented in the main text, we are foremost
interested in the quartic terms in $|{\bm \lambda}_i|$ and $\lambda_i^{\alpha}$, as these are the ones that lead to a selection 
between 1-${\mathbfit k}$ and 4-${\mathbfit k}$ states on which we henceforth focus. We substitute Eq. (\ref{chiexpan}) for $\chi^{\alpha\beta}_i$ and $ \chi^{\gamma\delta}_j$ in Eq. (\ref{TAP2}). The following subset of terms are found to potentially be able to differentiate between 1-${\mathbfit k}$ and 4-${\mathbfit k}$ states at the quartic order:
 \begin{equation}
 \label{finTAP1}
\Omega_1=-\beta\frac{1}{4(45)^2}\sum_{ij} |\bm \lambda_i|^2|\bm \lambda_j|^2\sum_{\gamma\delta}(J_{ij}^{\gamma\delta})^2, 
\end{equation}
\begin{equation}
\label{finTAP2}
\Omega_2 =-\beta\frac{1}{(45)^2} \sum_{ij}\sum_{\alpha\gamma\delta}J_{ij}^{\alpha\gamma}J_{ij}^{\alpha\delta} |\bm \lambda_i|^2\lambda_j^{\gamma}\lambda_j^{\delta} ,
\end{equation}
\begin{equation}
\label{finTAP3}
\Omega_3 =-\beta\frac{1}{(45)^2} \sum_{ij}\sum_{\alpha\beta\gamma\delta}J_{ij}^{\alpha\beta}J_{ij}^{\gamma\delta}\lambda_i^{\alpha}\lambda_i^{\gamma}\lambda_j^{\beta}\lambda_j^{\delta}.
\end{equation}

To calculate these terms, we used a computer program which directly calculates the sums by calculating 
$\bm \lambda_i$ for 1-$\bm k$ and 4-$\bm k$. $J_{ij}^{\mu\nu}$ is known through the Hamiltonian in Eq. (1) of the main text and the Ewald summation method has been considered to compute the contribution from the long-range dipolar interaction.
The reason these terms can distinguish between 1-$\mathbfit k$ and 4-$\mathbfit k$ is because they can not be reduced to onsite terms, an effect that happens with the lower order terms in the expansion with respect to $\beta$, i.e., the first two terms in Eq. (\ref{expansion1}), which constitute the s-MFT free energy. 
Fig. 2 in the main text presents the corrections to free energy beyond s-MFT 
arising from the computation of  $\sum_{i=1}^3(\Omega_i(\text{1-}\bm {k})-\Omega_i(\text{4-}\bm {k}))$.

\section{Monte Carlo Simulation Details}

To further investigate the 1-$\mathbfit{k}$ versus 4-$\mathbfit k$ selection mechanism in the Hamiltonian of Eq. (1) in the main text, we performed Monte Carlo simulations of three-component classical spins on the pyrochlore lattice. We used the parallel tempering\cite{ptempt} method to assist with thermal equilibration. The simulations were performed for three system sizes $L=4,6, 8$ where the system is composed of $L^3$ conventional cubic unit cells and each cell contains 16 lattice sites (pyrochlore structure). Even system sizes $L$ are required for the simulation cell to be commensurate with the 
$(\frac{1}{2}\frac{1}{2}\frac{1}{2})$ order.
To account for the long-range dipolar interaction, we employed the Ewald summation method \cite{Enjalran_MFT, MichelTTO}. For each temperature considered, about $10^8$ spin flips per spin were attempted while maintaining an average acceptance of approximately 50$\%$. 

As described in the main text, we define a \emph{distance parameter}, $d_{ik}, i=1,4$, in order to distinguish between the 1-$\mathbfit k$ and 4-$\mathbfit k$ states. To do so, we consider the 4-dimensional (4$D$) Euclidian space spanned by the four 1-$\mathbfit k$ states. In this 4$D$ space,  there are eight  points, that we denote as $\{\hat{p}_{1k}\}$, corresponding to 1-$\mathbfit k$ states (4 of them are defined in Table \ref{tab:conf1k} and 4 are the time-reversed of those), which can be represented as: 
\begin{equation}
 \hat{p}_{1k}=\left( \pm 1, 0, 0, 0\right), \\
 \left( 0,\pm 1, 0, 0\right), \\
 \left( 0,0,\pm 1, 0\right), \\
  \left( 0, 0, 0,\pm 1\right).
  \end{equation}
These points reside on the 4 axes of this 4$D$ space.  
Since the 4-$\mathbfit k$ states are linear combinations of the 1-$\mathbfit k$ states with coefficients $\hat{\psi}_a=\pm\frac{1}{2}$ (see Eq.(\ref{lincomb})), there are sixteen points corresponding to them 
\begin{equation}
\{\hat{p}_{4k}\}=\left(\pm\frac{1}{2}, \pm\frac{1}{2},\pm\frac{1}{2},\pm\frac{1}{2}\right).
\end{equation}
In the simulation, we measured $\psi_{a}$ defined in the main text and Eq. (\ref{eq2main}):
\begin{equation}\label{eq:mi}
\psi_{a}(\tau)=\frac{1}{N}\sum_{i}\mathbfit{S}_i(\tau)\cdot\hat{e}_{\mathbfit{k}_{a}}(\mathbfit{r}_i), 
\end{equation}
where $\psi_a(\tau)$ corresponds to the value of $\psi_a$ at the Monte Carlo step $\tau$ and $a=0, \cdots , 3$. $\hat{e}_{\mathbfit{k}_{a}}(\mathbfit{r}_i)$ is given by Eq. (\ref{eq:dir}). Consequently, the order parameter corresponding to $\{\frac{1}{2}\frac{1}{2}\frac{1}{2}\}$ ordering wave vectors, can be represented in the 4$D$ space as: 
\begin{equation}
|\bm\psi(\tau)|^2=\big\langle \sum_{a=0}^3\psi_a(\tau)^2\big\rangle_{\tau}.
\end{equation}
where $\langle\cdots\rangle_{\tau}$ represents Monte Carlo average.
Since we foremost care about the orientation of this vector in the 4$D$ space, once $|\bm\psi(\tau)|$ acquires a finite value at $T<T_c$, we define:
\begin{equation}
\mathbfit{\hat{\psi}}(\tau)=\frac{1}{\sqrt{\sum_{a=0}^3\psi_a^2}}(\psi_0(\tau), \psi_1(\tau), \psi_2(\tau), \psi_3(\tau)).
\end{equation}
from which we define:
\begin{eqnarray}
\tilde{d}_{1k/4k}(\tau)&\equiv&|{\mathbfit{\hat{\psi}}}(\tau)-\hat{p}_{1k/4k}|, 
\end{eqnarray}
At a particular temperature, at each $\tau$ instance, eight and sixteen values are obtained for 
$\tilde{d}_{1{k}}(\tau)$ and $\tilde{d}_{4{k}}(\tau)$, respectively. The minimum value among the eight/sixteen determines the $d^{\mathrm{min}}_{1{k}}(\tau)$ and $d^{\mathrm min}_{4{k}}(\tau)$. We then define  $ d_{1{k}/4{k}}$ at temperature $T$ as
\begin{equation}
d_{1{k}/4{k}} =\big\langle d^{\mathrm{min}}_{1{k}/4{k}}(\tau)\big\rangle_{\tau},
\end{equation}
which we refer to as the \emph{distance} parameter at the temperature considered. 
The results for $|\bm\psi|^2$ and $d_{1{k}/4{k}}$ are presented in Fig. 3 of the main text. We note that the low temperature distances show a different feature from the intermediate temperature regime. 

We close this section with the following observation.
Consider the four-dimensional unit vector $\hat{\psi}$. The 4-${\bm k}$ state corresponds to 
$\hat{\psi}^{(4)}=(1/2,1/2,1/2,1/2)$ while the 1-${\bm k}$ state has $\hat{\psi}^{(1)}=(1,0,0,0)$.
 It is straightforward algebra to show the 2-${\bm k}$ state, $\hat{\psi}=(1/\sqrt{2},1/\sqrt{2},0,0)$ 
has $d_{1k}=d_{4k}=\sqrt{2-\sqrt{2}}$. 
The low-temperature behavior of the $d_{i{\bm k}}$ in Fig. 3 of the main text may therefore suggests that the system is entering a 2-${\bm k}$ state at low-temperature.


\section{$(\frac{1}{2}\frac{1}{2}\frac{1}{2})$ and 1-${\bm k}$ vs 4-${\bm k}$ from second nearest-neighbor exchange}

Because they are often of considerable strength, we focused in the main body on the dipole-induced 
$(hhh)$ degeneracy-lifting into $(\frac{1}{2}\frac{1}{2}\frac{1}{2})$.
Yet, there are some rare-earth pyrochlore oxides, such as Yb$_2$Ti$_2$O$_7$ and Pr$_2$(Sn,Zr)$_2$O$_7$,
where the magnetic moment is sufficienly small that the dipolar interactions, ${\cal H}_{\rm dip}$,
 may not be the leading perturbation to ${\cal H}_0$. Rather, the second nearest-neighbor interactions, ${\cal H}_2$ might well be. Furthermore, it is of interest to consider the problem of the pyrochlore lattice with magnetic transition metal ions where dipolar interactions are definitely negligible compared to superexchange. We thus consider below the case $J_{\rm dip}=0$, but $J_2 \ne 0$.
Also, for completeness, we consider a different, but equivalent,
 representation of the anisotropic exchange couplings in ${\cal H}_0$ \cite{balents}.


The nearest-neighbor Hamiltonian, Eq. (1.a) in the main text, can be written in terms of local coordinates, ($lc$), and the couplings used in Ref. [\onlinecite{balents}]
\begin{eqnarray}
\label{eqross}
{\mathcal{H}}_0^{lc}&=&\sum_{\langle i,j \rangle}\{ J_{zz}S_i^zS_j^z-J_{\pm}(S_i^+S_j^- + S_i^-S_j^+ )\\ \nonumber
&&+J_{\pm\pm}(\gamma_{ij}S_i^+S_j^+ + \gamma_{ij}^{\ast}S_i^-S_j^- ) \\ \nonumber
&&+J_{z\pm}[S_i^z(\zeta_{ij}S_j^++\zeta_{ij}^{\ast}S_j^-) + i\leftrightarrow j]\},
\end{eqnarray}
where the $4\times 4$ matrix 
$\zeta_{ij}$ is given in Ref.[\onlinecite{balents}] and $\gamma=-\zeta^{\ast}$.


\begin{center}
 \begin{figure}
\includegraphics[width=9.5cm]{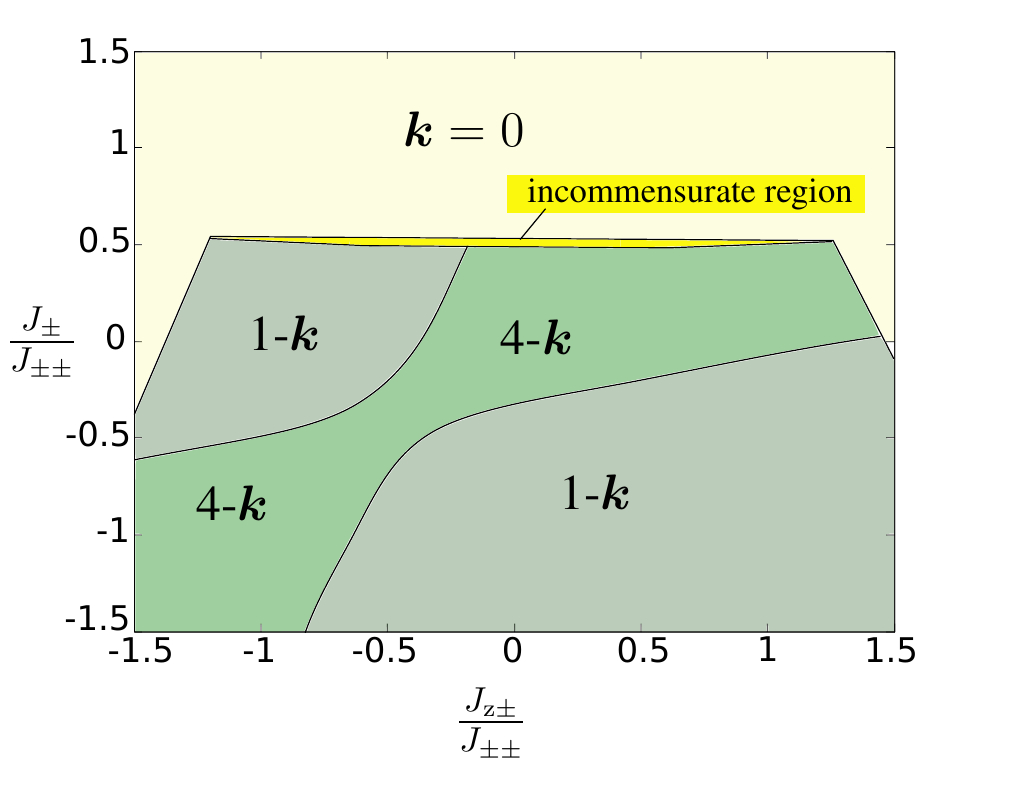}
\caption{ (Color online) Ordering wave vectors at $T=T_c$. The area denoted by 1-${\mathbfit k}$ and 4-${\mathbfit k}$ has $(\frac{1}{2}\frac{1}{2}\frac{1}{2})$ ordering wave vector. The first order E-TAP correction (Eq. (\ref{TAP2})), leads to the selection of either a 4-${\mathbfit k}$ or 1-${\mathbfit k}$ state (with corresponding labeled regions in the phase diagram depending whether the free energy is lower for 4-${\mathbfit k}$ or 1-${\mathbfit k}$ ordered phase). This phase diagram corresponds to ${\mathcal{H}}={\mathcal{H}}_0^{lc}+{\mathcal{H}}_{2}$ where ${J_{2}}/{J_{\pm\pm}}=-0.02$, $J_{zz}=0$ and  ${\mathcal{H}}_2=\sum_{\langle\langle i,j\rangle\rangle}J_2{\mathbfit S}_i\cdot{\mathbfit S}_j$.\label{fig: Wedge}}
\end{figure}  
\end{center}
 
This representation has appeared in a number of recent publications on pyrochlore magnets \cite{balents, columbic,ETO, sungbin}. In what follows, first, we consider only a nearest-neighbor Hamiltonian, ${\mathcal{H}}_0^{lc}$. 
We identify the critical modes at $T_c$ \cite{Enjalran_MFT}. These modes specify the ordering wave vector that first become soft (critical) as the temperature is decreased. 
As discussed in the main body of the paper, we obtain a line-degeneracy with ordering wave vectors $\{hhh\}$ with arbitrary $h$ in a certain region of the parameter space colored dark in Fig. (\ref{fig: Wedge}). Adding the beyond-nearest neighbor interactions such as ferromagnetic second- and/or antiferromagnetic third-nearest-neigbors and/or long-range dipolar interaction, lifts this degeneracy and favors $\{\frac{1}{2}\frac{1}{2}\frac{1}{2}\}$ as an ordering wave vector \cite{Matt1,cepas1,cepas2,Wills_Gd2Sn2O7}.

As stated above, we considered in the main text long-range dipole-dipole
 interactions as a source of the
selection of a $(\frac{1}{2}\frac{1}{2}\frac{1}{2})$ ordering wave vector. 
Here, we instead consider a ferromagnetic second nearest-neighbor interaction as perturbation.
We repeat all calculations leading to the computation of the TAP correction $\Omega$ in Eq. (\ref{TAP2}).
The results are illustrated in Fig. \ref{fig: Wedge}. In this case, we observe regions corresponding to $\mathbfit k=0$ and $\mathbfit k =(\frac{1}{2}\frac{1}{2}\frac{1}{2})$ ordering wave vectors. 
Using the  E-TAP method, we identify subregions corresponding to 1-$\mathbfit k$ and 4-$\mathbfit k$ states in the $\mathbfit k =(\frac{1}{2}\frac{1}{2}\frac{1}{2})$ region (see Fig. \ref{fig: Wedge} caption). The terms obtained in  Eqs. ({\ref {finTAP1}, \ref{finTAP2}, \ref{finTAP3}) can raise/lower the free energy of the 1-$\mathbfit k$ state with respect to 4-$\mathbfit k$ thus breaking the degeneracy.
 We also, observe a region in parameter space in which the ordering wave vector is incommensurate (specified in Fig. \ref{fig: Wedge}). The study of this region is beyond the scope of the present work.


 
 



\end{document}